# Transient Sulfate Aerosols as a Signature of Exoplanet Volcanism

Amit Misra[1,3], Joshua Krissansen-Totton[2,3], Matthew C. Koehler[2,3], Steven Sholes[2,3]

[1]Department of Astronomy, University of Washington, Seattle WA 98195, USA

[2]Department of Earth & Space Sciences, University of Washington, Seattle WA 98195, USA

[3]University of Washington Astrobiology Program, Seattle, WA 98195, USA

Keywords: Volcanism, Habitability, Biosignatures, Extrasolar Terrestrial Planets, Planetary Atmospheres

## Abstract

Geological activity is thought to be important for the origin of life and for maintaining planetary habitability. We show that transient sulfate aerosols could be a signature of exoplanet volcanism, and therefore a geologically active world. A detection of transient aerosols, if linked to volcanism, could thus aid in habitability evaluations of the exoplanet. On Earth, subduction-induced explosive eruptions inject $SO_2$ directly into the stratosphere, leading to the formation of sulfate aerosols with lifetimes of months to years. We demonstrate that the rapid increase and gradual decrease in sulfate aerosol loading associated with these eruptions may be detectable in transit transmission spectra with future large-aperture telescopes, such as the James Webb Space Telescope (JWST) and European Extremely-Large Telescope (E-ELT) for a planetary system at a distance of 10 pc, assuming an Earth-like atmosphere, bulk composition, and size. Specifically, we find that a S/N of 12.1 and 7.1 could be achieved with E-ELT (assuming photon-limited noise) for an Earth-analog orbiting a Sun-like star and M5V star, respectively, even without multiple transits binned together. We propose that the detection of this transient signal would strongly suggest an exoplanet volcanic eruption, if potential false positives such as dust storms or bolide impacts can be ruled out. Furthermore, because scenarios exist in which $O_2$ can form abiotically in the absence of volcanic activity, a detection of transient aerosols that can be linked to volcanism, along with a detection of $O_2$, would be a more robust biosignature than $O_2$ alone.

**Introduction**

Transiting exoplanets present a unique opportunity to identify and characterize potentially habitable worlds. As an exoplanet transits its host star, some of the star's light is absorbed and scattered in the planet's atmosphere, and the resulting transit transmission spectrum can be used to characterize the planet's atmosphere (e.g., Seager and Sasselov 2000; Brown 2001; Hubbard et al. 2001). This technique has been used to detect molecular absorption features for many Jupiter and Neptune-sized planets (e.g., Charbonneau et al. 2002; Vidal-Madjar et al. 2003; Barman et al. 2007; Pont et al. 2008; Deming et al. 2013; Fraine et al. 2014). Some planets show very weak absorption features, suggestive of a high-altitude cloud or haze layer (e.g., Sing et al. 2011; Sing et al. 2013; Kreidberg et al. 2014; Knutson et al. 2014a,b). With the upcoming launch of the James Webb Space Telescope (JWST) and the construction of larger ground-based telescopes such as the European Extremely Large Telescope (E-ELT), it may be possible to use transit transmission spectroscopy to characterize terrestrial planets in the near future (e.g., Deming et al. 2009; Kaltenegger and Traub 2009; Kaltenegger et al. 2010; Belu et al. 2011; Rauer 2011; Hedelt et al. 2013; Snellen et al. 2013; Misra et al. 2014a,b; Rodler and Lopez-Morales 2014). Here we examine transient sulfate aerosols ($H_2SO_4$ aerosols) in the transit transmission spectra of terrestrial extrasolar planets. We discuss how a detection of transient sulfate aerosols could be indicative of volcanism, and more broadly, a geologically active world.

Geologically active worlds in the habitable zone are likely more favorable environments for the emergence and persistence of life than geologically dead worlds in the habitable zone. Here we consider any volcanically active planet to be geologically active, and therefore geologically active planets can have plate tectonics or a stagnant lid. While geological activity is not necessary for habitability like liquid water is assumed to be, based on what is seen on Earth, planets without any form of geological activity are likely to be at a disadvantage for the origin and long-term maintenance of life. A de novo genesis of life appears to be reliant on chemical potential and redox gradients that were common in volcanic and hydrothermal settings during the Hadean and early Archean (Sleep et al. 2011; Stüeken et al. 2013 and references therein). Subaqueous hydrothermal vents powered by serpentinization of oceanic crust generated broad pH gradients (~6 for seawater to 9-11 for venting fluids) providing geochemically free energy for nascent life on early Earth (Sleep et al. 2011). Serpentinization also creates organic carbon molecules through Fischer-Tropsch-like reactions that may have been necessary for incipient biochemistry (Holm and Charlou 2001; Stüeken et al. 2013). Furthermore, modern life uses coupled redox pairs for electrochemical energy, and relies on "ion-tight" membranes and membrane enzymes to maintain disequilibrium (Mulkidjanian et al. 2012 and references therein). As the earliest organisms could not have had these complex membranes, one possibility is that environmental redox gradients, such as those that can be found in hydrothermal systems, could have supported protocells (Mulkidjanian et al. 2012; Stüeken et al. 2013). Conditions generated by geological activity were probably not only important for the origin of life but also for the subsequent radiation of the biosphere. Nisbet and Sleep (2001) postulate that if it were not for the perpetual supply of erodible rocks and steady introduction of volcanic cations, life would have slowly diminished due to a lack of bioessential elements. The overwhelming majority of electron donors and acceptors involved in early Earth ecosystems, and likely the origin of life, were sourced from subaerial and subaqueous volcanoes as well as hydrothermal systems (Canfield et al. 2006) that may have been commonly associated with plate boundaries. It should be noted that Earth-like plate tectonics might not be necessary for the recycling of bioessential elements and replenishment of redox gradients. Even in stagnant lid regimes, resurfacing via vertical transport can fulfill this role. For instance, Io is rapidly resurfaced by volcanism alone; the volume of lava flows and plume deposits are sufficient cover the entire surface in a 1m layer every century (Phillips et al. 2000).

Geological activity may help maintain long-term habitability via controlling climate stability. On Earth, the long-term carbonate-silicate cycle, a product of plate tectonics, has regulated the concentration of $CO_2$ in Earth's atmosphere throughout geologic time (Walker et al. 1981). Climate stability may also potentially be maintained in a stagnant lid regime (see for instance Pollack et al. 1987), but crucially, at least some volcanic outgassing is required to replenish the drawdown of $CO_2$ by silicate weathering and prevent the onset of a snowball.

Given the importance of geological activity for the origin and persistence of life, observational techniques capable of detecting volcanic activity could aid in evaluating the habitability of terrestrial exoplanets. Geological activity is not an absolute requirement for life in the same way that free energy and a liquid solvent are required; for instance, it is conceivable that the subsurface oceans of icy-moons sustain life from surface oxidants alone (McKay et al. 2008). However, extrasolar astrobiology is primarily focused on biospheres that affect the surface and atmosphere on global scales (Kasting and Traub 2010), as these are the habitable environments that can be observed and characterized. The most readily detectable exoplanet biospheres will be large-biomass, high metabolism surface biospheres that persist for geologically significant timescales. This is because small biospheres or biospheres not in contact with the atmosphere will have minimal influence on atmospheric composition, and because short-lived biospheres are unlikely to temporally overlap with our observations. Based on the points raised in the previous two paragraphs, we argue geological activity is necessary for the genesis and maintenance of these detectable biospheres. We propose that transient sulfate aerosols could be a spectral signature for volcanic activity and that the detection of transient sulfate aerosols would help inform habitability evaluations.

Currently, there is limited understanding of how we might determine whether exoplanets have active volcanism. Identifying volcanically produced gases is one proposed method to infer active volcanism on an exoplanet. Volcanic gas emissions are primarily composed of $H_2O$, $H_2$, $CO_2$, $CO$, $SO_2$, and $H_2S$. Of these, only $SO_2$ and $H_2S$ are known to be found in large concentrations almost exclusively as the result of volcanic activity. Jackson et al. (2008) suggested that highly tidally heated planets could have sulfur-rich atmospheres. This is similar to what is seen in the solar system, in which Jupiter's moon Io is tidally heated and erupts sulfur-bearing gases. For these planets, the detection of sulfur species (such as $SO_2$) could be indicative of active volcanism. Kaltenegger et al. (2010) investigated $SO_2$ as an atmospheric proxy for detecting moderate to high amounts of explosive volcanism on exoplanets. They found that $SO_2$ absorption from eruptions 100x larger than the 1991 Pinatubo eruption could be detectable (signal to noise ratio, S/N, >3) in primary (transit) or secondary eclipse with JWST for an exoplanet closer than 2.5 pc. Their results indicate that $SO_2$ will only be detectable as tracer for volcanism for very large, infrequent (once every >10,000 years) eruptions and only for planets orbiting the 6 closest stars, so the utility of $SO_2$ as a practical tracer for volcanism is limited. There are additional complications for $SO_2$ as a signature of volcanism. Hu et al. (2013) showed that for planets in the habitable zone, large surface fluxes of $H_2S$ and $SO_2$ lead to the formation of sulfur-based aerosols rather than detectable amounts of either gas. Furthermore, for atmospheres where $SO_2$ is very abundant, $SO_2$ is not an unambiguous signature of volcanism. The abundance of $SO_2$ in Venus' atmosphere is 6 orders of magnitude greater than that of Earth's, and the $SO_2$ shows order of magnitude variations (similar to what might be expected following a large volcanic eruption). While some early authors speculated that the variability of $SO_2$ was due to volcanic activity (Esposito 1984; Esposito et al. 1988), no evidence for volcanic activity has been found on Venus yet. Another explanation for the variability of $SO_2$ is dynamical global circulation of $SO_2$ (Marcq et al. 2013). Volcanic activity cannot be ruled out as the source of varying $SO_2$, but based on findings in the solar system $SO_2$ may not be a clear sign of volcanism.

We investigate the connection between temporal variations in stratospheric sulfate aerosols to active volcanism. On Earth, explosive volcanic eruptions (typically with a Volcanic Explosivity Index (VEI) >4) can inject $SO_2$ directly into the stratosphere, where it photochemically reacts with water to form sulfuric acid aerosols with lifetimes ranging from months to years. Stratospheric aerosols consequent of volcanism have been studied extensively in the Earth's atmosphere, typically with regards to their effect on climate forcing (Lacis et al. 1992; Sato et al. 1993), but here we examine their spectral effects. In the results section we show that these spectral effects would be detectable through observation of transiting terrestrial exoplanets. This paper only demonstrates the feasibility this technique using Earth as a case study; we model transit transmission spectra for an exoplanet with the same mass, radius, bulk composition, atmospheric composition and time-averaged insolation as Earth. However, the technique could be applicable to a broad range of terrestrial planet atmosphere types and host stars (see discussion). The discussion section also considers the possible connections between transient stratospheric aerosols and plate tectonics.

**Methods**

We generated transit transmission spectra of the Earth to test the detectability of transient stratospheric sulfate aerosols as a tracer for volcanism. We assumed Earth-like bulk composition, radius, mass and atmospheric composition. We placed the Earth-analog planets around a Sun-like star and an M5 star at the inner edge of the habitable zone, at which an Earth-analog planet will roughly be at the same temperature as the modern-day Earth. We assume an Earth-like atmosphere for the Earth-M5V case for simplicity, which does not account for slight differences in atmospheric composition and temperature structure predicted by self-consistent climate and photochemistry modeling (see Segura et al. 2005, for example). Furthermore, we assume the aerosol concentration will be at Earth-analog levels for various eruption settings, though in practice the aerosol concentration depends on the UV activity of the star (see Discussion). As sulfate aerosol abundance increases, the atmosphere becomes more opaque in transit transmission, resulting in higher effective absorbing radii. We used a radiative transfer model with standard pressure-temperature profiles and gas mixing ratio profiles, along with aerosol data retrieved from Earth-observing satellites, to generate the spectra. We also adopted different cloud schemes to cover a range of possible exoplanet cloud fractions and cloud altitudes. Our model and model inputs are described below in greater detail.

*Model*

Variations in stratospheric sulfate aerosols due to explosive volcanism should produce a transient increase in the transit transmission spectrum of an exoplanet atmosphere. We used Spectral Mapping Atmospheric Radiative Transfer-Transit Transmission (SMART-T) to generate transit transmission spectra of an Earth-analog exoplanet (Meadows and Crisp 1996; Crisp 1997; Misra et al. 2014b). SMART-T combines spectrally dependent atmospheric optical properties with a limb transmission model to generate transit transmission spectra. The model includes gas absorption, clouds and aerosols, refraction, Rayleigh scattering, and collisionally induced absorption. SMART, or models based on SMART, has been validated against a number of Earth-observing data sets (Robinson et al. 2011; Misra et al. 2014b; Robinson et al. 2014a) and is therefore an appropriate tool to model Earth-like exoplanets.

We modeled the Earth using the tropical, mid-latitude and subarctic reference models at different seasons from McClatchey et al. (1972). We used the prescribed pressure-temperature profile, water and ozone mixing ratio profiles and assumed that $CO_2$, $N_2O$, $CO$, $CH_4$, and $O_2$ are evenly mixed in all cases. We used the tropical atmosphere model for latitudes -23° < l < 23°, the mid-latitudes atmosphere for -60° < l < -23° and 23° < l < 60°, and the subarctic atmosphere for -90° < l < -60° and 60° < l < 90°.

*Aerosol Data*

We used data from the Stratospheric Aerosol and Gas Experiment (SAGE) II (McCormick 1987) and Optical Spectrograph and Infrared Imager System (OSIRIS) (Llewellyn et al. 2004) to model the aerosol optical depth over time. For background aerosol levels, we used an aerosol profile from McCormick et al. (1996) from 1979, which followed a period of low levels of volcanic activity and is therefore a reasonable approximation for background aerosol levels. We confirmed that this extinction profile matches background aerosol extinction profiles measured by SAGE II and OSIRIS. For small eruptions (total aerosol optical depth (AOD) $\ll$ 10x background levels), the extinction profile roughly scales by the aerosol optical depth without any change in the shape of the vertical profile. Therefore, we assumed the same vertical aerosol profile in the stratosphere but scaled the extinction up by the factor of increase in AOD. For larger eruptions with AOD > 10x background levels, we use the average aerosol vertical profiles retrieved by SAGE II and Osiris for each month and each latitude bin, with no approximations made. Figure 1 shows the 1 micron aerosol extinction profiles for the peak aerosol loading following a small (Sarychev-sized) and large (Pinatubo-sized) eruption compared to background aerosol levels.

To model the extinction from aerosol particles we used the Bohren and Huffman (1983) Mie scattering code to calculate wavelength-dependent absorption and scattering. For small eruptions, we assumed the stratospheric aerosols were 75% $H_2SO_4$ particles with a lognormal size distribution with an effective radius of 0.2 microns (from SAGE data) and an effective variance of 0.07 microns (Hansen and Travis 1974), though we find our results are not strongly dependent on the effective variance. We used the Palmer and Williams (1975) optical constants for 75% $H_2SO_4$ solutions to generate the Mie scattering output. We also generated wavelength-dependent optical properties for larger effective radii, which are applicable to large eruptions like Pinatubo. Larger eruptions lead to larger particle sizes (English et al. 2013), and for the Pinatubo eruption the maximum mean particle size is 0.6 microns.

For each month in the dataset, we generated a transmission spectrum for each of the six latitude bins. We averaged the aerosol optical depth over respective latitudinal bins (e.g. northern tropical latitudes) and generated a monthly spectrum over each bin using (a) the reference model atmosphere, (b) the retrieved aerosol extinction profile from SAGE II, and (c) an assumed cloud structure. We considered three cloud cases: (1) no clouds, (2) realistic clouds with an opaque cloud deck at 6 km with 100% cloud coverage, and an opaque cirrus cloud deck at 10 km with 25% cloud coverage, and (3) complete cloud coverage at the tropopause or at 10 km, whichever was greater. We include cases 1 and 3 to demonstrate the maximum possible effect of variations in cloud coverage on variations in the transit spectrum for an Earth-like atmosphere. For the realistic cloud coverage case, we adopted 100% cloud coverage at 6 km because even clouds that are considered optically thin in direct beam can be optically thick in transit transmission due to the increased path length in limb geometries. Other tropospheric aerosols, such as dust, pollution and products of biomass burning also contribute to tropospheric opacity and will generally be optically thick below 6 km (Winker et al. 2013). We note that along with clouds, refraction limits the altitudes that can be probed to altitudes >14 km for the Earth-Sun case and almost all altitudes for the Earth-M5V case (Betremeiux and Kaltenegger 2013, 2014; Misra et al. 2014), such that cloud variability has a lower effect on the spectra for an Earth-analog orbiting a Sun-like star than an M dwarf.

**Results**

We generated spectra for the Earth over time with realistic cloud coverage and aerosol loading as an analog for future exoplanet observations. Figure 2 shows spectra of the Earth for three different aerosol loadings: background aerosol levels, and levels corresponding to peak aerosol optical depth following the 1991 Mt. Pinatubo eruption and the 2009 Sarychev eruption with realistic clouds included, for an Earth-analog planet orbiting an M5V star. Details about the eruptions can be found in Table 1. We chose Sarychev to represent a small eruption and Pinatubo to represent a large eruption. These spectra show that the effects of volcanic eruptions on the Earth's spectrum are significant. The Sarychev eruption leads to changes in the spectra predominantly between 0.4 and 1.35 microns, while the Pinatubo eruption leads to changes at all wavelengths. The difference in the wavelengths affected is because the average aerosol particle size was 0.5 microns for the Pinatubo eruption, as opposed to 0.2 microns for the Sarychev eruption. The Pinatubo eruption also led to larger changes in the spectrum because of the increased aerosol loading.

Figure 3 shows the mean effective absorbing radius over time for the periods of the Pinatubo eruption and for all eruptions post-2000 for the three different cloud schemes we modeled. This figure shows that the mean effective radius of the atmosphere (measured as the mean spectral value between 0.685 and 1.40 microns, or over the $i$ and $j$ bands) increases as the aerosol optical depth (normalized to background levels) increases. This is because for higher aerosol optical depths, the atmosphere becomes more opaque and is absorbing a great quantity of light. Figure 3 also shows the difference in effective absorbing radius between the different cloud schemes. The difference in effective absorbing radius from the realistic cloud case is 1.5-2 km for the two more extreme cases. In contrast, the changes in absorption for the Sarychev and Pinatubo eruptions are 1.5 and 10 km, respectively. While the change due to the Sarychev eruption is similar to a theoretical change in cloud coverage, it is unlikely that the global cloud coverage would change by 100%. Based on data from MODIS (MODerate-resolution Imaging Spectroradiameter, http://modis.gsfc.nasa.gov, Salomonson et al. 1989) between 2000 and 2014, the maximum day-to-day variations for the globally-averaged cloud-top height over the past 14 years is <1 km (change in average pressure level from 0.62 to 0.69 bars). The maximum day-to-day variations in fractional cloud cover are from 62% to 74%. Therefore, a global change in cloud height of 10 km and cloud cover by 100% is extremely unlikely for the Earth. Therefore, the effect of the Pinatubo eruption on the transit transmission spectrum of the Earth is of sufficient magnitude to preclude any confusion with changes in cloud coverage, cloud height (up to the 10-15 km levels tested), or both.

Finally, we note that there is observational evidence to support our model results. Increased aerosol loading in Earth's atmosphere leads to darker and redder lunar eclipses (Keen 1983). Transit transmission spectra of exoplanets probe geometries similar to those seen in lunar eclipse; the passage of sunlight through Earth's limb (and subsequent reflection from the moon) is analogous to the passage of light through an exoplanet's limb during transit (Pallé et al. 2009).

Thus, increases in stratospheric aerosol concentration due to volcanism should lead to changes in the transit transmission spectrum of an exoplanet.

**Observing Eruptions**

*Required Time Resolution*

The lifetime of stratospheric aerosols in the Earth's atmosphere is on the order of months to years. This lifetime is set by the fallout rate of aerosols, and so this time scale should be constant (to first order) for Earth-analog exoplanets orbiting a range of spectral types. For smaller eruptions, the spectral effect of the aerosols will only last a few months, meaning that detecting an eruption would require time resolution on the order of months. In transit transmission we can only observe the spectrum of a planet once every exoplanet year. This means that Earth-like planets orbiting Sun-like stars will likely not be good candidates for observing smaller eruptions. The best candidates will be potentially habitable planets orbiting smaller M dwarf stars, which can have periods that are <1 month. For larger eruptions, longer timescales could be appropriate. The effect of the Pinatubo eruption on the Earth's spectrum lasted for ~4 years (see Figure 2b for exact globally-averaged aerosol optical depths over time). This means that detecting large eruptions should be possible for habitable planets orbiting Sun-like or smaller stars.

*Prospects for Detectability*

We used the James Webb Space Telescope (JWST) exposure time calculator (ETC) (Sosey et al. 2012) for the Near Infrared Spectrograph (NIRSPEC) in single prism mode and the European Extremely Large Telescope (E-ELT) ETC to estimate the detectability of stratospheric sulfate aerosols after an eruption. The instrumentation has not been specified for E-ELT, so we used the generic E-ELT spectroscopic ETC[1] (For more details on JWST, see Gardner et al. 2006, and for more details on E-ELT, see Gilmozzi and Spyromilio 2007). We do not consider limitations on observing time with E-ELT due to transits not overlapping with night times, but in practice this would reduce the number of transits that could be observed. We also ignore the potential need for comparison stars to control for instrument and atmospheric systematic errors for observations with E-ELT. This would likely increase the noise for each observation and could limit the number of potential targets due to the requirement of finding similarly bright comparison stars in the same field of view as the target star.

---

[1] https://www.eso.org/observing/etc/

We calculated the wavelength-dependent noise level obtainable with one transit for the case of a small (an analog Sarychev) eruption and a large (an analog Pinatubo) eruption orbiting Sun-like stars and M5V stars at a distance of 10 pc. To estimate detectability of volcanic eruptions on Earth-analog planets orbiting Sun-like stars, we used the given model stellar spectra (G2V Phoenix or Pickles models) and normalized the star's flux to have a *V* magnitude of 4.83. For estimating detectability for Earth-analog planets orbiting M5 stars, we used a Phoenix stellar model (Hauschildt et al. 1999) with an effective temperature of 2800 K, a log of surface gravity of 5, and a solar metallicity. We normalized the flux of the M5V star to 5.22E-14 erg/s/cm$^2$/Angstrom at 1 micron, which is equivalent to a flux of 174 mJy. The *V* magnitude of the M5 star is 14.4. The E-ELT ETC does not have an option for user-input spectra, so we calculated the estimated noise at each wavelength by converting the model stellar flux to mJy and running the ETC simulation. We assumed a transit duration of 13 hr and an orbital period of 1 year for the Sun-like case, and a transit duration of 1.58 hr and an orbital period of 11.5 days for the M5V case. The noise levels are the inverse of the S/N returned from the exposure time calculators at each wavelength, multiplied by factors of $10^6$ (to convert the noise to ppm) and 2. There noise needs to be multiplied by $\sqrt{2}$ to account for the need to difference between in-transit and out-of-transit spectra, and another factor of $\sqrt{2}$ to account for comparison two different transit transmission spectra, one for background aerosol levels and one for post-eruption aerosols levels.

Small, Sarychev-sized eruptions may be at the threshold of detectability for E-ELT, but will not be detectable with JWST. The signal levels (defined as the difference between the post-eruption spectra and the background aerosol levels spectra) at 1 micron (with refraction included) for a small eruption for an Earth-like planet around a Sun-like star and an M5 star are 0.02 ppm and 0.74 ppm, respectively. However, the signal from stratospheric aerosols covers a very wide wavelength range. In Figure 3 we've averaged the change in effective radius over a wide wavelength range. For E-ELT, we selected 70 wavelength bins between 0.68 and 1.38 microns in which atmospheric transmission is fairly large, allowing for high-precision ground-based observations. Shortward of 0.68 microns, M dwarfs have very low flux levels (and therefore large noise levels), and longward of ~1.35 microns, there is a strong water absorption band that severely limits ground-based observations. For JWST, we selected all wavelength bins between 0.6 microns and 5.0 microns, which cover the full wavelength range of the NIRSPEC instrument in single prism mode. Assuming photon-limited noise (which should be appropriate for the bright stars that are likely targets for transit transmission spectroscopy), integrating the S/N over multiple wavelengths should be possible. Roughly, if $N$ wavelength bins are integrated over, the noise will decrease by a factor of $\sqrt{N}$ if the noise is photon-limited. The overall S/N for a small eruption is much less than 1 with JWST. The S/N is 1.3 for the Earth-Sun case and 0.9 for the Earth-M5 case for observations with E-ELT. For the M dwarf planet, because the orbital period is relatively short (11.5 days), it could be possible to average over 9 transits (a total of 3.25 months of time) to get to a S/N of ~3, which could be enough for a detection. Nevertheless, the spectroscopic effects of the 2009 Sarychev eruption would be at the threshold of detectability of E-ELT.

A Pinatubo-style eruption should be detectable for an Earth-like planet at a distance of 10 pc, perhaps even with JWST. The S/N at the height of the Pinatubo eruption would be 2.1 and 1.8 with JWST and 12.1 and 7.1 with E-ELT for the Sun-like and M5 case, respectively, even without any transits co-added. The effect of a Pinatubo-magnitude eruption lasts for ~4 years, so increasing S/N by co-adding transits should be feasible. For example, for an Earth-analog orbiting a Sun-like star, if the peak of the eruption was observed, the S/N for that 1 transit with JWST would be 2.1. If a second transit were observed one year later, the S/N for the second transit would be ~1.2. The third and fourth transit would have S/Ns of 0.7 and 0.4, respectively. Integrating over all transits would yield a best case S/N of 2.6. For a 'worst case' scenario, in which the peak is not observed, the total S/N for the Earth-analog orbiting a Sun-like star would be ~2.0. For an Earth-analog orbiting an M dwarf, multiple transits could potentially be observed each month, greatly increasing the S/N. There are over 130 potential transits for the M dwarf case over a 4 year period, and by integrating over all possible transit the total S/N with JWST for an Earth-analog orbiting an M5 star would be 11.8. Even if only one transit per month could be observed, the total S/N would be 7.3 with JWST. Similar binning over transits could be done with E-ELT observations as well. Therefore, if we observe an Earth-like exoplanet that is undergoing a Pinatubo-magnitude eruption we could detect the effects of that eruption with E-ELT and possibly JWST.

While JWST and E-ELT may be able to detect transient sulfate aerosols for planets within 10 pc, larger telescopes would be required to detect this effect for planets further away. There are 67 stars within 5 pc, of which 50 are M dwarfs and 10 are F, G, and/or K stars (Cantrell et al. 2013). An Earth-analog orbiting a Sun-like star has a ~0.5% probability of transiting, and an Earth-analog orbiting an M dwarf has a ~2% probability of transiting, meaning there is likely only 1 potentially habitable transiting planet within 5 pc, if every star hosts a potentially habitable planet. Multiple groups have estimated the occurrence rate of habitable zone (HZ) planets orbiting M dwarf to be near 0.5 (Bonfils et al. 2013; Gaidos 2013; Kopparapu et al. 2013), which would result in an average of 0.5 potentially habitable planets within 5 pc. There are 322 main sequence stars out to 10 pc, of which 248 are M dwarfs (Henry et al. 2006), increasing the number of potential habitable transiting planets out to 10 pc to ~8 (or 4 assuming 0.5 HZ planets per star). Thus, the applicability of the technique described here to JWST and E-ELT will likely be limited to a small number of targets. However, this method would be applicable to more stars than the Kaltenegger et al. (2010) method. Their results show that $SO_2$ could be detectable for an eruption that emits 100x more $SO_2$ into the atmosphere than a Pinatubo-sized eruption for the 6 closest stars (within 2.45 pc) for transit transmission observations, among which there is a very low probability of having even one transiting, potentially habitable planet. Furthermore, the Kaltenegger et al. (2010) calculations assume that it is possible for a large amount of $SO_2$ to build up in an Earth-like atmosphere, which has been shown to be implausible by Hu et al. (2013). They show that large abundances of $SO_2$ and $H_2S$ lead to aerosol formation, and therefore are not predicted to build up to detectable levels. However, the presence of an aerosol layer alone is not predictive of volcanism on an exoplanet. This leaves transient sulfate aerosols as the most detectable signature of exovolcanism on Earth-analog exoplanets.

This method will likely be more applicable to the next generation of ground and space-based telescopes after JWST and E-ELT. Within 30 pc, there would be ~100 ($4x3^3$, accounting for increase in volume) potentially habitable targets, greatly increasing the likelihood of a detection of transient sulfate aerosols. However, a detection at a distance of 30 pc is most likely not feasible with JWST or E-ELT, meaning that this technique may be most applicable to future large aperture space and ground-based telescopes. Given that observations of a star at 30 pc collect $1/9^{th}$ the light of a star at 10 pc, detecting transient sulfate aerosols for a planet at 30 pc would require a telescope with nearly an order of magnitude greater collecting area than the next generation of space- and ground-based telescopes.

**Discussion**

*Potential False Positives and False Negatives*

We have shown that a detection of a transient aerosol event may be possible with E-ELT and possibly JWST for Earth-analog exoplanets. However, transient aerosol detection alone would not be conclusive evidence of volcanism because there exist potential false positives, which we discuss below. A detection of transient aerosols would need to be combined with other spectral features (which may require additional observing time) and detailed modeling before a transient aerosol event could be confidently linked with active volcanism. It is also important to note that an absence of a transient aerosol signal is not indicative of an absence of active volcanism. Below we discuss scenarios in which even explosive eruptions may not result in a detectable transient aerosol signal. Regardless, transient aerosol signals (or lack thereof) would need to be evaluated within the planetary context before any conclusions can be drawn regarding geological activity on the exoplanet.

Martian dust storms exhibit similar patterns of sharp increases in aerosol levels, followed by a gradual decrease with timescales ranging from months to years (Pollack et al. 1979). Dust storms could thus cause a potential false positive for detecting changes in stratospheric sulfate aerosols. Silicate absorption features near 9.7 and 3.4 microns could be used to distinguish between a dust storm and an increase in sulfate aerosols (Grishko and Duley 2002). Additionally, on Earth, dust storms do not reach the stratosphere at the level that sulfate aerosols do, so it may be that large planet-wide dust storms like those we see on Mars require a thin, low pressure atmosphere.

If silicate absorption features will be used to distinguish between dust storms and explosive eruptions, then silica ash ejected into the atmosphere during eruptions must be ruled out as a false negative. Generally, volcanic ash rapidly (within days) falls out of the atmosphere, and does not become as geographically dispersed as the associated $H_2SO_4$ aerosols (see Schneider et al. 2012; Karagulian et al. 2010; Prata and Kerkmann 2007). Therefore, the effect of ash on the spectrum should be minimal and short-lived and will not significantly alter the results. Even in the unlikely scenario of an observation being made immediately following an eruption, when ash levels are high, and even if the ash cloud were on the planet's terminator (a requirement for having an effect on the transit transmission spectrum), the geographical extent would be small, extending to a few degrees in latitude and longitude at most, meaning that its effect on the spectrum would be very small.

However, this rapid fallout does not include particles that are smaller than 1 micron (Schneider et al. 2012). On Earth, these smaller particles (a very small percent of the total ejected ash) can accompany $SO_2$ into the stratosphere and have residence times that depend on particle size, shape, and density, as well as medium viscosity and turbulence (Fuchs 1989). It is conceivable that an exoplanet with a different atmosphere and mineralogy than Earth could host an explosive eruption that causes optical and chemical changes in the stratosphere from both sulfate aerosols and volcanic ash, given an abnormally large amount of very small ash particles. If this is the case, it might still be possible to distinguish between dust storms and ash-laden explosive eruptions. A dust storm should see a uniform transient signal across both sulfate aerosol and silicate absorption features, whereas a volcanic eruption are likely to introduce a greater amount of long-residence $SO_2$ into the stratosphere than silica ash, and so the absorption features of both sulfate aerosols and silica ash may reflect this disproportionally.

Bolides can explode in an atmosphere or impact a planetary surface, and generate stratospheric sulfate aerosol signals that can mimic those created by explosive volcanic eruptions. For example, Gorkavyi et al. (2013) detected and traced an aerosol excess belt in the stratosphere following the Chelyabinsk meteor (18 m diameter) impact in early 2013. This event was small in terms of changes in stratospheric aerosol levels as the aerosol belt was thin, and aerosols returned to background levels within a few months. Kring et al. (1996) demonstrate that carbonaceous chondrites containing up to 6% sulfur by weight, with diameters greater than 300 m, will generate vapor plumes that deposit amounts of sulfur in the stratosphere analogous to the 1883 Krakatau, 1963 Agung, and 1982 El Chichón eruptions (see Table 1 for details of each eruption). On Earth, these impacts are estimated to occur once every 10,000 years (Sigurdsson 1990; Kring et al. 1996). To estimate the relative frequency of such impacts and how they would compare to the frequency of explosive eruptions, it would require knowing parameters such as the age of the system, system dynamics, and initial reservoir of debris. Modeling of these parameters would be necessary to determine whether a transient aerosol signal is likely caused by an explosive volcanic eruption.

Planets for which sulfur-bearing species were injected at an altitude below optically-thick aerosol layers would likely be false negatives using this method. For example, explosive volcanism on Venus or Titan, if such existed, would most likely not significantly affect the global cloud and haze layers by a sufficient amount unless the sulfate aerosols were being produced high above the existing cloud and haze layers. Venus is optically thick in transit transmission to 90 km (Ehrenreich et al. 2012, Garcia-Muñoz and Millis 2012), so a transient aerosols signal would require an injection of sulfur more than 90 km above the surface. Titan is optically thick above 100 km in transit transmission (Robinson et al.

2014b), and would face a similar problem as Venus. Overall, transient sulfate aerosols will likely only be detectable for eruptions for which the plume height is greater than both the tropopause height and the height of an optically thick cloud layer. On Earth, eruptions with plume heights greater than 10-15 km should lead to transient aerosols signals. The plume height for the Pinatubo eruption is 40 km, so Pinatubo-like eruptions could be detectable for planets with tropopause heights and cloud heights less than 40 km.

Venus may also present a false positive because there are temporal variations in the cloud height and aerosol extinction whose cause is currently unknown. The cloud base altitude can vary by several kilometers, but only at high latitudes (Ignatiev et al. 2009). Additionally, the aerosol extinction in solar occultations at an altitude of 80 km varied by nearly a factor of a few in the near-infrared between 2006 and 2010 (Wilquet et al. 2012). For reference, a factor of 5 change in aerosol extinction would at most translate to a change in effective radius in transit transmission of 4 km. However, the aerosol extinction variability does not occur simultaneously over all altitudes, so the effect of these variations on the global transit transmission spectrum of Venus would be smaller. When averaged over the entire wavelength range considered here, the transient aerosol signal would be further reduced, though spectral modeling of Venus in transit transmission that incorporates these recent observations would be required to better estimate the maximum expected transient aerosol signal. Overall, Venus does exhibit transient aerosols, though at a much lower level than the predicted effect of a large explosive volcanic eruption like Pinatubo on an Earth-analog exoplanet.

The stellar UV flux incident on the planet can lead to a false negative for UV quiet stars and a potential false positive for active, flaring stars. The formation of sulfate aerosols requires OH or $O_2$, which are typically formed via the photolysis of $H_2O$ and $CO_2$, respectively, in the absence of biology. Photochemical simulations of aerosol formation for planets orbiting UV quiet stars show that aerosol formation can be decreased by orders of magnitude. Thus, transient sulfate aerosols may be difficult to observe for planets orbiting stars with very low levels of UV flux. In contrast, it is possible that stellar flares or other events that temporarily increase the stellar UV flux can lead to a transient increase in aerosol optical depth. This would require an abundance of either $H_2S$ or $SO_2$ in the planetary atmosphere, which would react with photochemically produced OH or $O_2$ after the flare to form sulfate aerosols. In this scenario, the star would have to have low levels of UV flux prior to the flare. To rule out such a case, one could look for evidence of a flare, which affect wavelengths beyond the UV (Kowalski et al. 2013). Furthermore, a transient sulfate aerosol signal that was triggered by a flare would still provide evidence for a flux of sulfur compounds into the atmosphere, presumably from volcanism. Photochemical modeling of sulfate aerosol formation for different stellar types and different UV fluxes, while beyond the scope of this present work, should help determine if changes in the incident UV radiation on a planet can lead to transient sulfate aerosol events that are not caused by volcanism.

Stellar activity could be another potential false positive. Starspots are cooler and fainter than the rest of the stellar disk, and can therefore affect the stellar flux. Starspots can also vary between transits, creating the inter-transit variation that would be seen during a transient sulfate aerosols event. Unocculted starspots could increase the measured transit depth, and occulted starspots could decrease it. To avoid confusion with stellar activity such as starspots, a detection of transient sulfate aerosols would likely need to occur over multiple transits. If a planet transited a starspot during one transit but not during the subsequent transit, a change in the effective radius of the planet would be observed, similar to what we predict for a planet undergoing a transient sulfate aerosol event. However, the starspot pattern on a star typically leads to changes in the light curve that are sinusoidal (Queloz et al. 2009), and would most likely not lead to the characteristic rapid rise and gradual decline in effective radius that we see for transient sulfate aerosols.

*Aerosol Formation in Anoxic Atmospheres*

While our calculations have been done for oxygen-rich, Earth-like atmospheres, sulfate aerosols form in anoxic atmospheres as well. Hu et al. (2013) modeled $H_2SO_4$ and $S_8$ aerosol formation in anoxic $H_2$, $CO_2$ and $N_2$-domianted atmospheres over a range of $H_2S$ and $SO_2$ volcanic fluxes. They found that $H_2$ and $CO_2$-domianted atmospheres preferentially form $S_8$ and $H_2SO_4$ aerosols, respectively, across all $H_2S$ and $SO_2$ flux levels. $N_2$-domianted atmospheres form $H_2SO_4$ if the ratio of $H_2S$ to $SO_2$ gases emitted via volcanism is lower than three times the present day ratio. For oxic atmospheres, even trace amounts of $O_2$ ($pO_2 > 10^{-5}$ PAL) result in the formation of $H_2SO_4$ over $S_8$ (Pavlov and Kasting 2002; Zahnle et al. 2006). Regardless, for any atmosphere with a surface sulfur flux, sulfur aerosols (either $H_2SO_4$ or $S_8$) are expected to form photochemically and fall out over an extended period of time, though more detailed photochemical modeling is needed to quantify the detectability and duration of volcanic events under different redox regimes. Therefore, this method should be applicable to oxidized and reduced atmospheres, as well as the oxic Earth-like atmospheres we have examined here.

*Aerosol signal over all spectral types*

We've presented results for transient sulfate aerosols for Earth-like planets orbiting a Sun-like star and an M5V star, but the signal should be detectable over a wide range of spectral types. The transient aerosol signal (i.e., the ppm difference between the post-eruption spectrum and the background aerosol spectrum) will scale with $R_*^{-2}$, where $R_*$ is the stellar radius, because the transmission signal is inversely proportional to the area of the projected stellar disk. The noise over one transit will scale with $(L_* \tau)^{-1/2}$, where $L_*$ is the stellar luminosity and $\tau$ is the transit duration. Qualitatively, the noise is proportional to the square root of the number of photons collected, which is proportional to the stellar luminosity and the integration time, which for one transit equals the transit duration. Using stellar parameters for F, G, and K stars (Zombeck 1990) and for M dwarfs (Reid and Hawley 2005) we find that the average S/N varies by less than a factor of 4 over all stellar types examined for one transit, with greater S/N for planets orbiting earlier type stars, which have longer transit durations. The average S/N over an equal integration time varies by less than 30% over all stellar types. As shown above, the S/N could be increased for planets orbiting later type stars by binning over multiple transits, which should be possible because the orbital periods are often <1 month. Therefore, while our results are calculated for the Sun-like and M5V cases, the S/N estimates should be comparable over all F, G, K, and M stars.

*Detectability Caveats*

The detectability calculations we've presented here represent an estimate of the S/N for detecting a transient aerosol event for an Earth-analog planet (Earth radius, mass, atmospheric composition) at a distance of 10 pc, assuming only photon-limited noise. More accurate estimates of detectability could be obtained by applying Bayesian retrieval techniques to model spectra with noise. Such models have been applied to exoplanet data (Madhusudhan et al. 2011; Benneke and Seager 2012; Line et al. 2013). In these models, the atmospheric properties are retrieved from a data spectrum and from posterior assumptions about the atmosphere. The parameters for the model atmosphere are sampled via a Markov Chain Monte Carlo. Integrating transient aerosol events into a retrieval model would provide more rigorous estimates for the detectability of these events.

As discussed previously, a detection of a transient aerosol signal is not necessarily evidence for geological activity. Further characterization of the planetary environment, either from detection of other spectral absorption features or from additional modeling work, would be needed before a claim of a detection of volcanism, and therefore geological activity could be made. Placing a spectral absorption feature within the planetary context is necessary for most absorption features, not just transient aerosols. For example, as mentioned previously, $SO_2$ is not an unambiguous signature of volcanism. And as we will discuss in a following subsection, $O_2$ is only a robust biosignature when considered in context. Likewise, a transient aerosol signal could only be linked to volcanism if potential false positives can be ruled out. Therefore, the detectability results presented here should be considered limits for detecting transient aerosols, not for conclusively detecting exovolcanism.

*Would transient sulfate aerosols imply active plate tectonics?*

Transient sulfate aerosols in the lower stratosphere are derived from explosive volcanic eruptions that advect sulfur material through the troposphere, avoiding rapid fallout. The detection of such transient aerosols, when false positives can be ruled out, provides compelling evidence for active volcanism, which alone would be an important finding in an exoplanet survey. Additionally, on Earth these explosive volcanic eruptions are normally the result of plate tectonics. The explosivity of an eruption is largely dependent on the silica and volatile (primarily $H_2O$ and $CO_2$) content of the magma. On Earth, silicate and volatile-rich magmas are typically generated in subduction zones under continental and island-arc settings and therefore the resulting stratovolcanism is highly explosive. In contrast, eruptions not associated with subduction zones, such as hot spot volcanism, are typically less explosive due to the low silica and volatile content of the magma. Consequently, almost all explosive volcanic eruptions large enough to markedly increase stratospheric aerosol optical depth occur in subduction zone settings (as illustrated in Figure 4). Therefore these transient aerosol events may also be indicative of an active tectonic regime, or at the very least a geologically active planet.

There have been numerous discussions of the planetary properties, namely observables such as mass and radius, required to initiate and sustain plate tectonics (Valencia et al. 2007a,b; Valencia and O'Connell 2009, O'Neill and Lenardic 2007). However, Lenardic and Crowley (2012) show that the ability for a planet to enter into a tectonic regime depends primarily on the geological history of the planet. Multiple modes of tectonic regimes (e.g. mobile plates, stagnant lid, and episodic tectonics) can occupy the same parameter space making it difficult to predict whether these extrasolar planets will be favorable for plate tectonics if only the bulk physical properties are known (e.g. mass, radius, composition, etc.). Therefore, it is helpful to have an additional constraint on the probability of a planet having plate tectonics.

It is conceivable, however, that terrestrial exoplanets with differing bulk compositions or with large initial volatile reservoirs, could produce explosive volcanic eruptions that are not associated with plate tectonics. For instance there is compelling evidence for explosive volcanism on Mars despite there being no evidence for active, Earth-like plate tectonics (Mouginis-Mark et al. 1982; Wilson and Head 1994; Kerber et al. 2012). On Mars, basaltic explosive eruptions are relatively common for two reasons. Firstly, Mars' lower gravity and low atmospheric pressure enhance gas exsolution and the fragmentation of magma during ascent to the surface, thereby leading to more explosive eruptions (Wilson and Head 1994). Secondly, the interaction between subsurface water and magma increases magma volatile content and therefore eruption explosivity (Wilson and Head 1994). For exoplanets, surface gravity can be constrained from the exoplanet mass and radius and lower limits on atmospheric pressure can be set by detections of absorption from dimer molecules, whose absorption is more sensitive to density and pressure than typical rotational and vibrational molecular absorption features (Misra et al. 2014a). This could help to inform the likelihood that observed explosive volcanism is associated with plate tectonics. Although there is no way of remotely observing subsurface water content, it has been argued that on Mars there was a transition from explosive to effusive volcanism around 3.5 Ga attributed to declining subsurface water content (Robbins et al. 2011).

Evidently, more work needs to be done on the influences of planetary composition and evolution on eruption explosivity. The future detection of transient sulfate aerosols in an exoplanet atmosphere must be carefully considered in the context of other observations before any tentative inference to plate tectonics can be made. However, in instances where observations of a stellar system can rule out false positives, the detection of transient sulfate aerosols would be strongly indicative of explosive volcanism and therefore a geologically active world. As discussed previously, geological activity is crucial for the long-term maintenance of a highly productive (and therefore remotely detectable) biosphere. Consequently, even if a definitive inference to plate tectonics can never be made, the mere detection of explosive volcanism on an exoplanet merits further observation of the exoplanet.

*Sulfate-aerosols and oxygenic biosignatures*

It has long been argued that the remote detection of atmospheric oxygen/ozone in a terrestrial exoplanet's atmosphere would constitute a promising biosignature (Lippincott et al. 1967; Walker 1977; Angel et al. 1986). On Earth, oxygen production via photosynthesis is 700 times larger than abiotic oxygen production via water photolysis and hydrogen escape (Kasting 1997). In the absence of photosynthesis, Earth's steady-state oxygen column abundance would be many orders of magnitude smaller than present atmospheric levels and not detectable remotely (Kasting 1997). However, it is disputed whether pathological scenarios exist whereby abiotic oxygen/ozone could build up on a habitable zone Earth-like planet and be mistaken for biogenic oxygen. Photochemical modeling by Segura et al. (2007) suggests that even in high-UV (more photodissociation) environments, abiotic oxygen could not build up to detectable levels, regardless of volcanic outgassing rates. Hu et al. (2012) and Harman et al. (in preparation) were generally able to reproduce these results with independent photochemical models, although the former study reported that detectable abiotic oxygen could accumulate in very high pCO$_2$ atmospheres. In contrast, Domagal-Goldman (2014), Tian et al. (2014), and Wordsworth and Pierrehumbert (2014) identified various scenarios whereby abiotic oxygen could readily build up to detectable levels on Earth-like planets. Suffice to say the question of oxygen false-positives is an unresolved theoretical issue.

Nevertheless, even if atmospheric oxygen is a potentially ambiguous biosignature, the detection of transient sulfate aerosols in the presence of an oxygen-rich atmosphere could help resolve the ambiguity between biotic and abiotic sources of oxygen. The detection of transient sulfate aerosols presumably implies a source flux of reduced volcanic gases into the observed exoplanet's atmosphere (exempting unlikely scenarios involving oxidizing volcanic gases, the outgassing of redox-neutral species such as SO$_2$ and H$_2$O will always be accompanied by outgassing of reduced species such as H$_2$S and H$_2$). Hu et al. (2012) demonstrate that for very high CO$_2$ (1 bar) atmospheres, abiotic oxygen can only accumulate to detectable levels if there is no volcanic outgassing. If volcanic outgassing is comparable to the modern-Earth levels, then the column abundance of both O$_2$ and O$_3$ drops dramatically, thereby eliminating the oxygen false positive. Domagal-Goldman et al. (2014) also show that for a wide variety of stellar types and CO$_2$-abundances, the column abundance of abiotic oxygen/ozone drops dramatically as the volcanic outgassing rate is increased. Evidently the simultaneous detection of transient sulfate aerosols and atmospheric oxygen could strengthen the case for biogenic oxygen. For any given exoplanet, photochemical modeling of its atmosphere, stellar spectrum observations and careful consideration of stellar and orbital history will be required to evaluate how well the inference to biology is strengthened. Unfortunately it is unclear whether the oxygen source flux could be quantified by such modeling; the observation of a single episode of volcanism would not constrain the time integrated flux or redox-state of the planet's volcanic outgassing. Perhaps if the planet were observed continuously for several years and multiple volcanic episodes were detected this would put a lower bound on volcanic outgassing rates, and by extension a lower bound on the oxygen source flux (explosive volcanism is a small fraction of the total volcanic outgassing on Earth).

**Conclusions**

Transient aerosol loading in terrestrial exoplanet stratospheres may be detectable in transit transmission spectra of exoplanets within 10 pc using future telescopes such as JWST and E-ELT. We have shown that eruptions as large as the 1991 Mt. Pinatubo eruption would be detectable with E-ELT with a S/N>7 for Earth-analog planets for most F, G, K, and M stars even without multiple transits binned over. A Sarychev-sized eruption could be at the threshold of detectability (S/N ~3) with E-ELT if multiple transits could be binned over.

The detection of such transient sulfate aerosols would be suggestive of extrasolar volcanic activity, which would make that planet a more promising candidate for follow-up observations since volcanism is important for the origin and maintenance of life. However, a confirmation of volcanic activity may require additional modeling and perhaps further spectroscopic observations to rule out potential false positives such as dust storms and bolide impacts. Transient sulfate aerosols could also be suggestive of active plate tectonics since on Earth, the explosive eruptions that inject sulfate aerosol-forming $SO_2$ into the stratosphere occur almost exclusively at convergent plate boundaries. However, even a tentative claim of plate tectonics would require a detection of transient sulfate aerosols coupled with other system/planetary observations and modeling.

Transient aerosols in the presence of atmospheric $O_2$ would also constitute a more robust biosignature than that detection of $O_2$ alone if the transient aerosols can be linked to volcanism. In the absence of surface fluxes of reduced gases, oxygen can potentially build up in an atmosphere even without the presence of life, making $O_2$ alone an ambiguous biosignature. However the detection of volcanism would strongly suggest a source of oxygen-consuming reduced gases and would thus strengthen the case for biogenic oxygen.

**Acknowledgments**

We thank John Baross, Eva Stüeken, Roger Buick and David Catling for helpful comments. We thank the three anonymous reviewers for their comments that greatly improved the paper.

This work was supported in part by the NASA Astrobiology Institute's Virtual Planetary Laboratory Lead Team, funded through the NASA Astrobiology Institute under solicitation NNH12ZDA002C and Cooperative Agreement Number NNA13AA93A.

Analyses and visualizations used in this study were produced with the Giovanni online data system, developed and maintained by the NASA GES DISC.

**Author Disclosure Statement**

No competing financial interests exist.

Table 1: Eruption Parameters.

| **Eruption** | **Date** | **VEI** | **SO$_2$ (Tg)** | **Plume Height** |
|---|---|---|---|---|
| Sarychev | June 2009 | 4 | 1.2 (Haywood et al. 2010) | ~14 km (SVERT[2]) |
| Pinatubo | June 1991 | 6 | 20 (Bluth et al. 1992) | >40 km (Oswalt et al. 1996) |
| El Chichón | Mar-Apr 1992 | 5 | 8 (Thomas et al.1983) | ~17 km (Varekamp et al. 1984) |
| Krakatoa | Aug 1883 | 6 | 2.9 (Devine et al. 1984) | >40km (Self and Rampino 1982) |
| Agung | Feb 1963 | 5 | 2.5-7 (Self and King 1996) | >20km (Self and King 1996) |

---

[2] SVERT: Sakhalin Volcanic Eruption Response Team (Documented by the Smithsonian Institute/USGS weekly volcanic activity report) (http://www.volcano.si.edu)

**Figures**

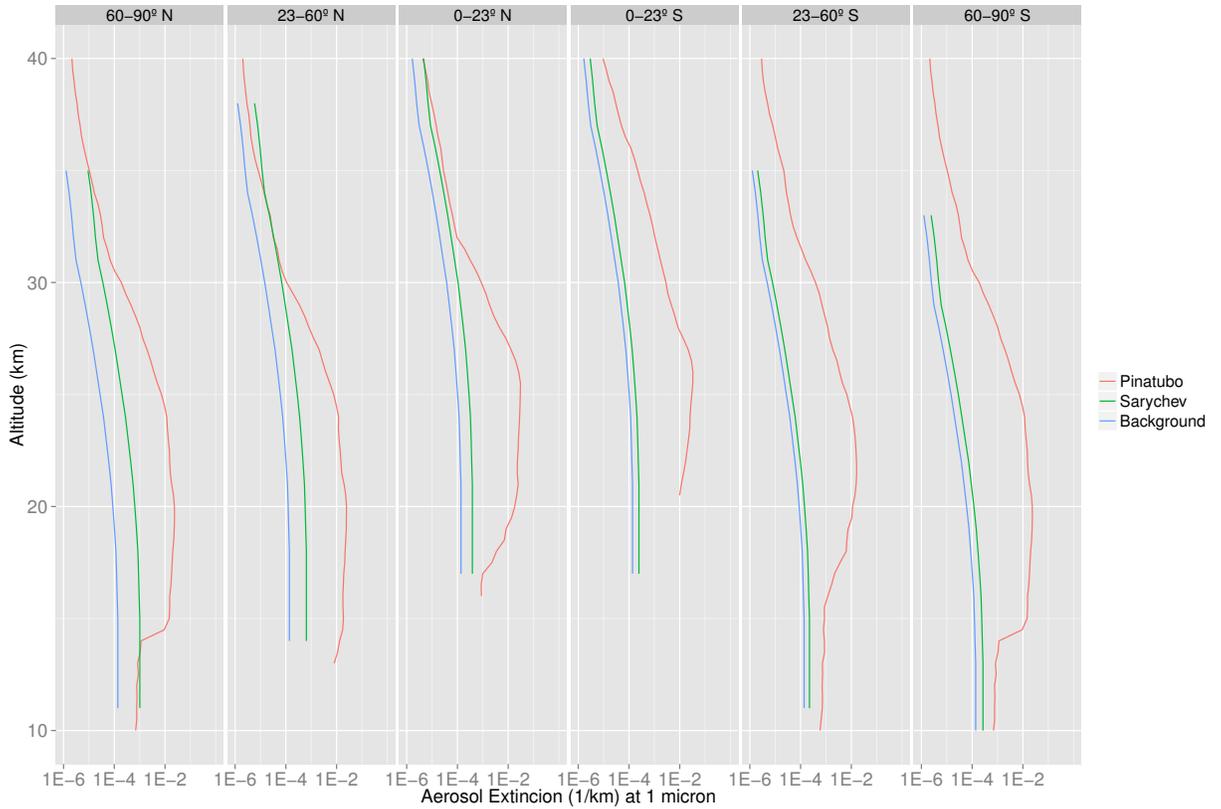

Figure 1: H$_2$SO$_4$ aerosol extinction profiles following the Sarychev and Pinatubo eruptions, with background aerosol levels shown for comparison. The peak aerosol extinction levels were 1 month after the Sarychev eruption and 6 months after the Pinatubo eruption. The background aerosol profiles are from McCormick et al. (1996) and closely match the retrieved profiles from SAGE II and Osiris for periods with low aerosol loadings. The Sarychev profiles shown are scaled up

from the background profiles by the average aerosol optical depth measured as a function of latitude. Based on retrieved profiles from SAGE II and Osiris, this approximation is valid for eruptions that result in increases in aerosol optical depth (or extinction) by less than a factor of 10. The Pinatubo aerosol extinction profiles are the retrieved profiles with no approximations made.

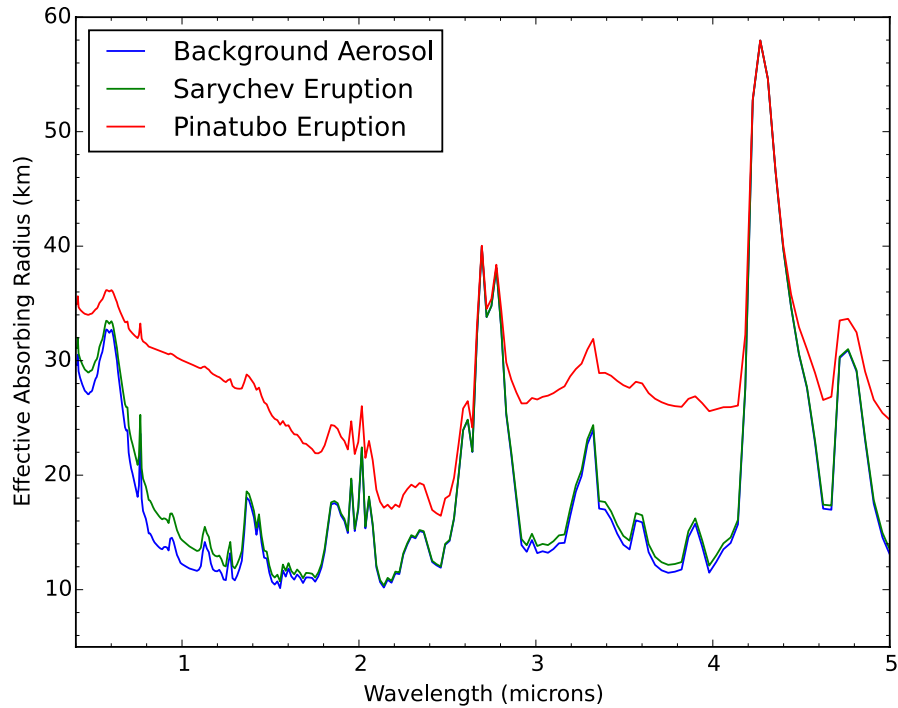

Figure 2: Spectra (in effective absorbing radius of the atmosphere) of the Earth during different amounts of aerosol loading: with background aerosol levels, after the 2009 Sarychev eruption and after the eruption of Mt. Pinatubo in 1991. The peak aerosol levels following each eruption are shown. The aerosol lifetimes following each eruption are roughly 3 months for Sarychev and 4 years for Pinatubo. The spectra are shown for the period after the eruption when the aerosol loading was greatest. While the effect of increased aerosol loading on the Earth's spectrum is small for the Sarychev eruption, it is large (>20 km) for the Pinatubo eruption. The wavelength cutoffs are 0.4 microns (at which noise levels increase for ground-based observations) and 5.0 microns (the longward cutoff for JWST NIRSPEC). If multiple transits can be binned over, we find that it may be possible to detect (S/N >3) a Pinatubo-sized eruption with JWST or a Sarychev-sized eruption with E-ELT for an Earth-analog planet orbiting an M5 star. Assuming photon-limited noise, a Pinatubo-sized eruption for an Earth-analog planet at a distance of 10 pc may be detectable with E-ELT with S/N >7 for both the Sun-like and M5V case, even without multiple transits binned over.

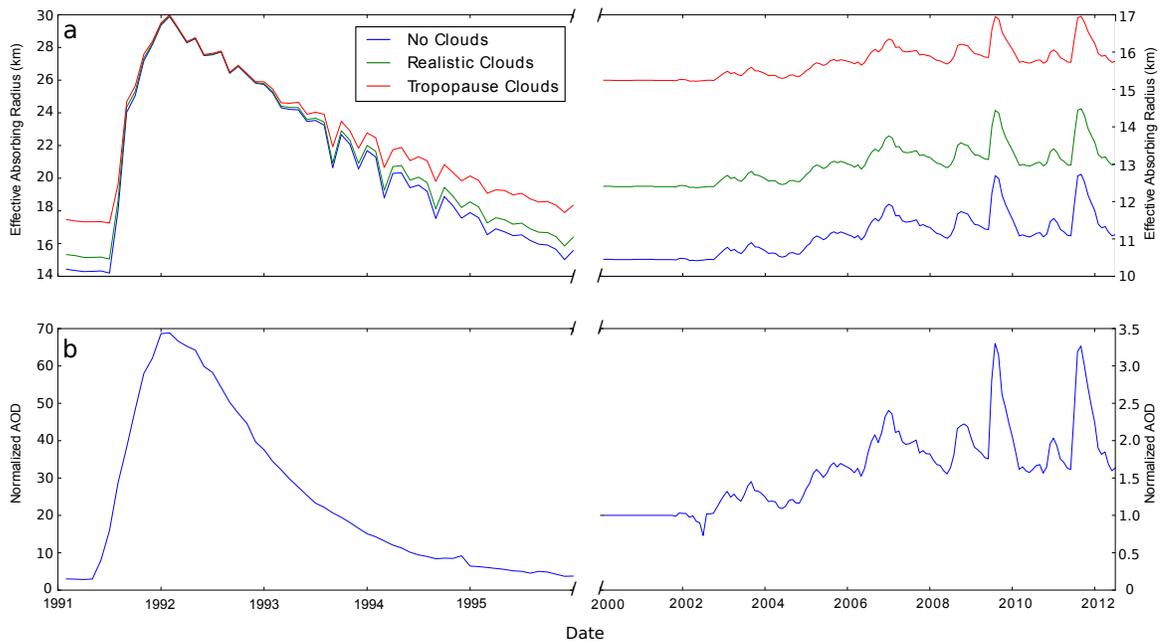

Figure 3: a) Mean effective radius for different cloud types and b) aerosol optical depth (normalized to background aerosol levels in early 2000's) over time for an Earth-analog planet orbiting an M5V star. Note the different y-axis scales for the left and right hand sides. The mean effective radius of the atmosphere increases as aerosol optical depth increases. This figure also shows that the change in effective radius between different cloud schemes is smaller than the change due to volcanic eruptions for a large eruption like Pinatubo. The change due to a Sarychev-type eruption is roughly equal to 100% changes in cloud coverage between the realistic case and either of the end-member cases. The 100% change in cloud coverage results will be even smaller for an Earth-analog planet orbiting a Sun-like star because refraction limits the altitudes that can be probed to >14 km, while for an

Earth-analog planet orbiting an M5V star the altitudes that can be probed are >1 km.

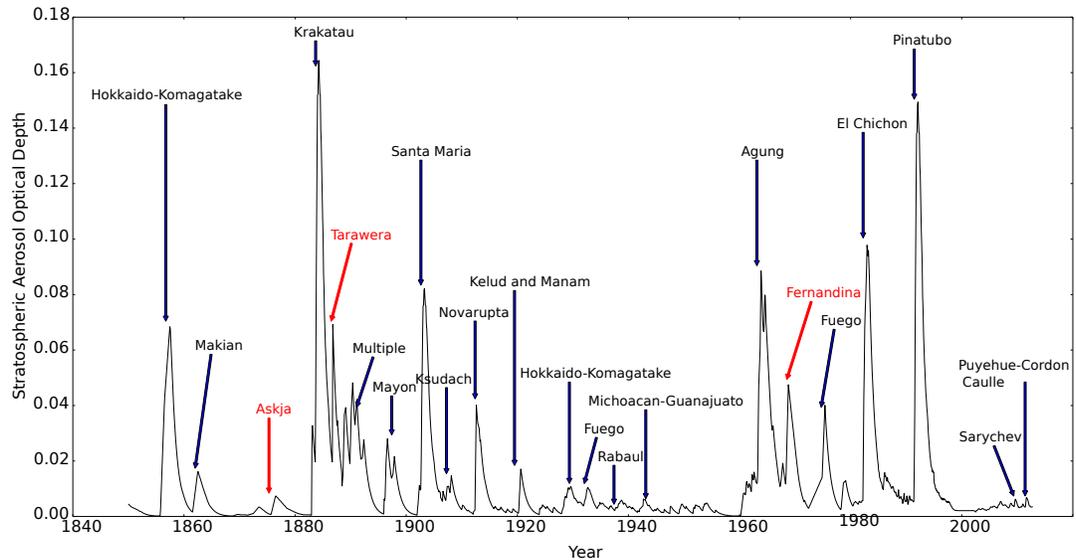

Figure 4: Globally averaged stratospheric aerosol optical depth for the last 160 years is plotted in black. Selected explosive eruptions (VEI of 4, 5 or 6) are denoted by arrows together with the optical depth excursions with which they coincide. Most of the named eruptions occurred at subduction zone settings; the three exceptions are highlighted in red. The 1886 Tarawera eruption was a rare basaltic Plinian eruption (Walker et al. 1984), albeit associated with a back-arc basin (Lau-Harve-Taupo). The 1968 Fernandina eruption was associated with the Galapagos hotspot, although the tectonic setting is complicated by the intersection

of the Nazca, Coscos and Pacific plate boundaries. The 1875 Askja eruption in Iceland also occurred at a complex tectonic setting (Geiger et al. 2010). In general, we see that increased stratospheric aerosol optical depth is caused by eruptions in subduction zone settings, and even the few exceptions are not entirely disconnected from tectonic activity. Volcanic eruptions were obtained from the Smithsonian Institute Global Volcanism Program database (http://www.volcano.si.edu/) and stratospheric optical depth was obtained from NASA GISS (http://data.giss.nasa.gov/modelforce/strataer/).